\documentclass{nature}
\usepackage{multicol}
\usepackage{graphicx}
\usepackage{textcomp} 
\usepackage{ifthen} 
\usepackage{xifthen} 
\usepackage{color} 
\usepackage{soul} 
\usepackage{amsmath} 
\usepackage{amssymb} 
\usepackage{amsfonts}
\usepackage{lineno}
\usepackage{pdfpages}
\usepackage{siunitx} 
\usepackage{hyperref} 
\usepackage{tikz} 
\usetikzlibrary{shapes}

\makeatletter
\let\saved@includegraphics\includegraphics
\AtBeginDocument{\let\includegraphics\saved@includegraphics}
\renewenvironment*{figure}{\@float{figure}}{\end@float}
\makeatother

\newcommand{\fig}[2][]{%
\ifthenelse{\isempty{#1}}
{Fig.~\ref{#2}}
{Fig.~\ref{#2}#1}
}
\newcommand{\Fig}[2][]{%
\ifthenelse{\isempty{#1}}
{Figure~\ref{#2}}
{Figure~\ref{#2}#1}
}
\author{J. Doster$^1$, S. Hoenl$^1$\footnote{Present adress: IBM Research – Zurich, S{\"a}umerstrasse 4, CH-8803 R{\"u}schlikon, Switzerland}, H. Lorenz$^2$, P. Paulitschke$^2$ \& E.\,M. Weig$^{1}$}

\begin{document}
\title{Strong Strain-Induced Coupling between Nanomechanical Pillar Resonators}
\maketitle
\begin{affiliations}
	\item Department of Physics, University of Konstanz, Universit{\"a}tsstraße 10, 78457 Konstanz, Germany
	\item Fakult{\"a}t f{\"u}r Physik and Center for NanoScience (CeNS), Ludwig-Maximilians-Universit{\"a}t, Geschwister-Scholl-Platz 1, 80539
M{\"u}nchen, Germany
\end{affiliations}
\begin{abstract} 
Networks of coupled resonators are an ubiquitous concept in physics, forming the basis of synchronization phenomena,~\cite{Pikovsky.2001} metamaterial formation,~\cite{Engheta.2006} nonreciprocal behavior~\cite{Caloz.2016} and topological effects.~\cite{Hasan.2010} Such systems are typically explored using optical or microwave resonators.~\cite{Ozawa.2018} In recent years, mechanical resonators have entered the stage as universal building block for resonator networks, both for their well-controlled mechanical properties and for their eigenfrequencies conveniently located in the radio-frequency regime.
Vertically oriented nanomechanical pillar resonators~\cite{Paulitschke.2013} are ideally suited for the dense integration into large resonator networks. However, to realize the potential of these promising systems, an intrinsic coupling mechanism needs to be established. Here, we demonstrate strain-induced, strong coupling between two adjacent nanomechanical pillar resonators. The coupling is mediated through the strain distribution in the joint substrate caused by the flexural vibration of the pillars, such that the coupling strength can be controlled by the geometric properties of the nanopillars as well as their separation. Both, mode hybridization and the formation of an avoided level crossing in the response of the nanopillar pair are experimentally observed. The coupling mechanism is readily scalable to large arrays of nanopillars, enabling all-mechanical resonator networks for the investigation of a broad range of collective dynamical phenomena.
\end{abstract}

In recent years, arrays of coupled macroscopic mechanical resonators have been employed to demonstrate topologically protected transport of mechanical excitations.~\cite{Susstrunk.2015,Nash.2015,He.2016,SerraGarcia.2018}
In principle, the underlying concept can be readily transferred to nanoscale implementations,~\cite{Peano.2015,Fosel.2017} offering the advantage of straightforward on-chip integration. However, sufficiently strong nearest-neighbor coupling is difficult to achieve with nanomechanical resonators, and nanoscale topologically protected transport has to date only been achieved using phononic crystal architectures~\cite{Cha.20180627} or parametric coupling.~\cite{Tian.20180826} Similarly, nanomechanical implementations of nonreciprocal signal transduction through resonator arrays rely on parametric coupling~\cite{Huang.2016} or optomechanical interactions,~\cite{Fang.2017,Bernier.2017,Peterson.2017} and the synchronization of nanomechanical resonators is either based on external feedback~\cite{Matheny.2014} or mediated by an optical radiation field.~\cite{Zhang.2015}

In contrast, intrinsic mechanical coupling between adjacent nanomechanical beam or string resonators has been reported and relies on the strain distribution in a shared clamping point.~\cite{Karabalin.2009,Gajo.2017,Pernpeintner.2018} In a related approach, the membranes constituting the building blocks of nanomechanical phonon waveguides exchange energy through physical connections.~\cite{Hatanaka.2014,Cha.2018}

Here, we translate the concept of strain-induced intrinsic coupling to vertically oriented nanomechanical pillar resonators sharing the same substrate.~\cite{Yeo.2014,Anguiano.2017,Lepinay.2017,Rossi.2017} We consider a pair of inverted conical nanopillar resonators like the one displayed in~\fig[a]{Figure1}. The pillars are etched into a (100) GaAs substrate using reactive ion etching~\cite{Paulitschke.2013} and feature eigenfrequencies in the range of a few megahertz. 
The fundamental flexural eigenmode of each of the pillars exhibits two orthogonal polarization directions with slightly different eigenfequencies as a result of fabrication imperfections.~\cite{Rossi.2017}
With respect to the indistinguishable \textless100\textgreater\ crystal directions, these modes will in the following be referred to as 'horizontal' (H) and 'vertical' (V) polarization (see Fig.~S1).
In total the pillar pair has four eigenmodes, which are correspondingly labelled LH, LV, RH, and RV for the left (L) and right (R) pillar, respectively.

Measurements are performed in vacuum $<\SI{e-4}{\milli\bar}$ and at room temperature using piezo-actuation,
while the response of the nanopillars is read-out by scanning electron microscope imaging~\cite{Janchen.2002} or by optical detection,~\cite{Yeo.2014} focusing a laser with $\lambda=\SI{635}{\nano\metre}$ wavelength onto the head of one nanopillar and detecting the vibration-induced modulation of its reflection.

To evaluate the stress distribution in the substrate, we perform finite element simulations as displayed in \fig[b and c]{Figure1}. \Fig[b]{Figure1} suggests a significant stress overlap between the two pillars, indicating that the structural features in the interjacent area will affect their coupling. 
Therefore, the geometry of this part of our model reflects the realistic pillar geometry and features a narrowing mesh as highlighted in \fig[c]{Figure1}. Overlapping stress profiles are found for $d\lesssim 4r$. The pillar separations of the investigated samples have been chosen accordingly.

We first investigate a nanopillar pair with bottom radius $r\approx\SI{310}{\nano\metre}$, height $H\approx\SI{7}{\micro\metre}$, and center-to-center distance $d\approx\SI{1.3}{\micro\metre}$. The taper angle of $\SI{1.1}{\degree}$ is the same for all nanopillars discussed in this work. The nanopillar pair is driven at frequency $f_{\text{drive}}$ in a scanning electron microscope, allowing to image the resulting envelope of its mechanical vibration.
In total, we find four well separated vibrational modes with no spectral overlap (see Fig.~S1 of Supplementary Information), which are identified with the eigenmodes of the pillar pair.
One of them (LV) is shown in \fig[a and b]{Figure2}, which display the vibrational envelope of the nanopillar pair driven at $f_{\text{drive}}=f_{\text{LV}}$, imaged from the top and in a tilted view. A more careful inspection of the mode reveals that in \fig[a and b]{Figure2} not only the left resonator vibrates with a large amplitude but also the right resonator exhibits a simultaneous vibration, albeit with a much smaller amplitude (see also movie M1 and M2).
This is also apparent from \fig[c]{Figure2} which shows the vibrational amplitude of both pillars extracted from micrographs obtained at different drive frequencies $f_{\text{drive}}$ in the vicinity of the resonance $f_{\text{LV}}$. Clearly, the two amplitudes evolve simultaneously with the drive frequency $f_{\text{drive}}$, which indicates the existence of a hybridized mode arising from coupling between the two pillars mediated by the joint substrate.

To obtain a more thorough understanding of the observed inter-pillar coupling, we measure the response of a second pillar pair with $r\approx\SI{310}{\nano\metre}$, $H\approx\SI{7}{\micro\metre}$, and $d\approx\SI{1}{\micro\metre}$ using the optical detection setup. Thermal tuning, readily implemented by the laser used for optical detection, is employed to sweep the eigenfrequency of the higher-frequency pillar through that of the other pillar which remains largely unaffected (see Supplementary Information for details). \Fig{Figure3} shows an avoided crossing of a pair of nanopillar resonators, which is indicative of strong mechanical coupling. A fit of the data using the model of two linearly coupled harmonic oscillators is also included as a black solid line.~\cite{Novotny.2010} It yields a coupling rate $g/2\pi=\SI{8.3\pm1.8}{\kilo\hertz}$ which exceeds the linewidth of the mechanical resonances $\Delta f\approx\SI{3.5}{\kilo\hertz}$. Hence, we demonstrate strong intrinsic coupling between the two nanopillar resonators, for this specific set of geometry parameters.
The two modes to the left of the avoided crossing are assigned to the vertical vibration of each resonator (LV, RV) via scanning electron micrographs shown in the insets of \fig{Figure3}, where the tuned pillar with the higher frequency corresponds to the right pillar. Further evidence for the inter-pillar coupling arises from the evolution of the vibration amplitudes in \fig{Figure3}. Since the laser is focused on the frequency-tuned right pillar, mainly the vibration of this one resonator is detected and the vibration of the left pillar is only weakly resolved through the stray field of the laser. Clearly, the transition of the strong signal of the right pillar from the upper to the lower branch of the avoided crossing reflects the hybridization of the two pillars near their resonance at $\SI{425}{\micro\watt}$ and $f_r=\SI{7.401}{\mega\hertz}$. 

The observed strong coupling supports the conclusions drawn from \fig{Figure2}, the data for which was acquired without frequency tuning the right pillar and with variable $f_{\text{drive}}$. An avoided crossing measured for this pair with $d=\SI{1.32}{\micro\metre}$ is included in Fig.~S1. The untuned situation probed in the scanning electron microscope measurements corresponds to the state of the system towards the leftmost edge of the avoided crossing: Far from resonance, the individual eigenmodes of the two pillars dominate the response, however a slight hybridization is already apparent as a consequence of their coupling.

Strain-mediated coupling through the substrate is expected to depend on geometrical parameters of the pillar pair. In particular, the bottom radius $r$ of the nanopillars, their height $H$ and their center-to-center distance $d$ have proven influential.
In the following, we investigate the dependence of the nanopillar coupling strength on these geometry parameters in measurements and finite element simulations. Coupling rates are obtained from  fits to the measured avoided crossings as described above. Finite element simulations (\fig[b and c]{Figure1}) are performed to evaluate the eigenfrequencies of the two nanopillars. Frequency tuning is incorporated by a variation of Young's Modulus $E$ of one of the nanopillars, mimicking thermal tuning. The level splitting $g/2\pi$ is then also obtained by fitting the model of the coupled oscillators to the simulation data (see Supplementary Information). We investigate the dependence of $g/2\pi$ on each of the parameters $d$, $r$ and $H$ while the other two parameters remain fixed. Only when sweeping the bottom radius $r$, $d$ is adapted accordingly to ensure a constant edge-to-edge distance.

\Fig[a]{Figure4} shows the measured level splittings of vertical modes  and \fig[b, c and d]{Figure4} display the simulation results for the three different sweep parameters. The simulations in \fig[b]{Figure4} show increasing coupling when the center-to-center distance of the two nanopillars is decreased. This is a typical signature of strain coupling,~\cite{Ovartchaiyapong.2014} and in agreement with the measurements in \fig[a]{Figure4}. Furthermore, our simulations predict an increasing coupling strength with the bottom radius of the two nanopillars (\fig[c]{Figure4}), as a consequence of the strain caused at the pillar foot upon its deflection, and the experimental data in \fig[a]{Figure4} clearly reflects this behavior. Finally, the simulations show a decrease of the coupling strength with the height of the pillars (\fig[d]{Figure4}), again in consequence of the resulting strain profile. The dependence on the height, however, can not clearly be established in \fig[a]{Figure4} since the differences in coupling for the two investigated pillar heights are smaller than the error bars.

The results in \fig[a]{Figure4} have been obtained for pillar pairs with different orientations on the substrate as indicated in the inset of \fig[a]{Figure4} (see supplementary Fig. S4 for details). It is expected that the angular dependence of Young's modulus of the crystalline GaAs substrate should lead to an angular dependence of the strength of the strain-induced coupling (see supplementary Fig.~S4). However, differences in coupling strength with pillar orientation are not experimentally resolved, likely because of the fabrication-induced disorder in the pillar geometry governing the uncertainty in our measurements. To validate the influence of the substrate's Young's Modulus, further studies should address the vibration polarization of hybridized (symmetric or antisymmetric) modes for arbitrary orientation of the pillar pair.

In conclusion, we reveal strain-induced coupling between two adjacent nanomechanical pillar resonators in the strong coupling regime, and show mode hybridization as well as the formation of an avoided level crossing under thermal tuning of one of the nanopillars. The coupling strength is found to depend on the center-to-center distance of the nanopillars, as well as their diameter and height. Numerical finite element simulations reproduce the observed scaling, confirming that the coupling between two nanopillar resonators is mediated by strain in the substrate which acts as a shared clamping point.

Vertically oriented nanopillars are ideally suited for the integration into large arrays. Uncoupled arrays of nanopillars are employed for phonon guiding or cellular force tracking.~\cite{Paulitschke.2018} In addition, III-V semiconductor-based nanopillars offer the possibility of integrating optically active quantum dots,~\cite{Yeo.2014} enabling addressing and readout via their optical properties.
 The demonstrated coupling mechanism is readily scalable to large arrays of nanopillars with geometry-controlled nearest neighbor coupling. This opens the door to two-dimensional all-mechanical resonator networks for the investigation of a broad range of collective dynamical phenomena, including topological effects,~\cite{Salerno.2017,Yang.2015,Fleury.2016,Fosel.2017}
 and may pave the way towards nanomechanical computing~\cite{Mahboob.2008} or nanomechanical implementations of neural networks.~\cite{Vodenicarevic.2017}
\nolinenumbers
\begin{addendum}
 \item Financial support from the European Unions Horizon 2020 programme for Research and Innovation under grant agreement No. 732894 (FET Proactive HOT), the Deutsche Forschungsgemeinschaft via the Collaborative Research Center SFB 767, the Volkswagen Foundation through grant Az I/85 099, as well as the German  Federal  Ministry  of  Education  and  Research through contract no. 13N14777 funded within the European QuantERA cofund project QuaSeRT and the programm ``Validation of the Technological Innovation Potential of Scientific Research - VIP'' is gratefully acknowledged.
 \item[Competing Interest]The authors declare no competing interests.
 \item[Author Contributions] Measurements were performed by J.D., while sample fabrication was done by J.D., S.H., H.L., and P.P.; measurement results and data interpretation were discussed by all authors.
\end{addendum}
\bibliography{PillarPairs_1}
\newpage
\begin{figure}
\includegraphics[width=\columnwidth/2]{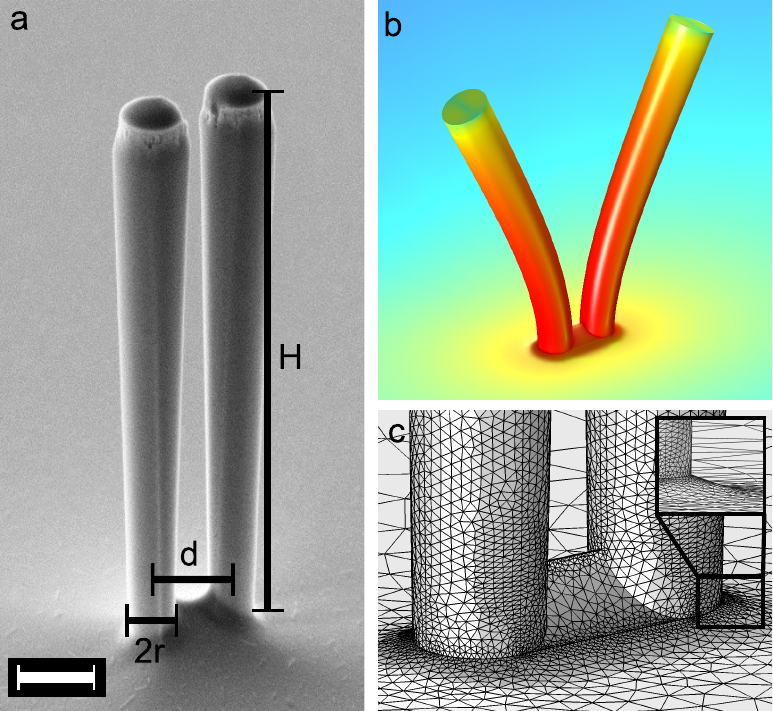}
\caption{\label{Figure1}\bf Nanomechanical System. a\sf, Scanning electron micrograph of a pair of nanopillar resonators with a height of $H\approx\SI{7}{\micro\metre}$, a center-to-center distance of $d\approx\SI{1.1}{\micro\metre}$ and a foot radius of $r\approx\SI{310}{\nano\metre}$ in a $\SI{60}{\degree}$ tilted view. The taper angle of $\SI{1.1}{\degree}$ gives rise to a somewhat larger head radius. Scale bar corresponds to $\SI{1}{\micro\metre}$. \bf b\sf, Finite element simulation of the stress distribution between two (identical) oscillating nanopillar resonators. On the color scale, red (blue) corresponds to high (low) stress, respectively. \bf c\sf, Mesh of the finite element simulation model narrowing at the pillar foot allowing for a more careful analysis of the clamping point as the most relevant region for the coupling. The inset shows a zoom of the transition from a nanopillar to the substrate.
}
\end{figure}
\begin{figure}
\includegraphics[width=\columnwidth/2]{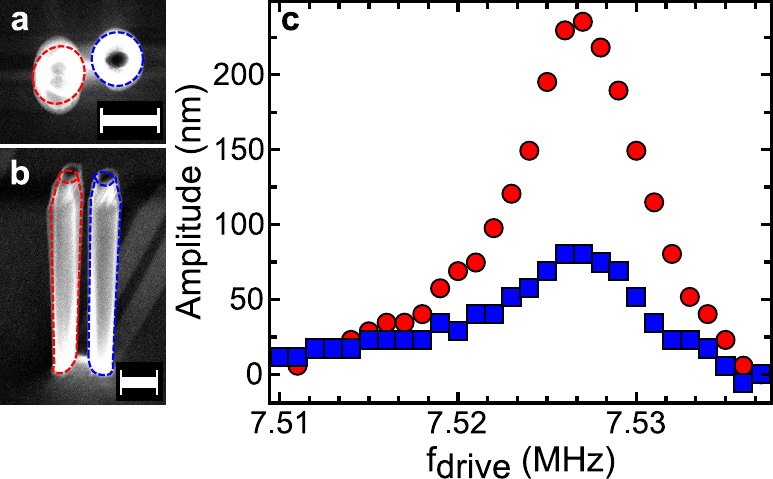}
\caption[]{\label{Figure2}\bf Hybridized mode. a\sf, Scanning electron micrograph of a resonantly driven ($f_\text{LV}=\SI{7.527}{\mega\hertz}$) pair of nanopillars ($H\approx\SI{7}{\micro\metre}$, $r\approx\SI{310}{\nano\metre}$, $d\approx\SI{1.3}{\micro\metre}$) imaged from the top and \bf b, \sf in a $\SI{60}{\degree}$ tilted view. Red and blue dotted lines indicate the circumference of the undriven left and right resonator, respectively. \bf c\sf, Amplitude for different drive frequencies $f_{\text{drive}}\approx f_{\text{LV}}$ of left (\tikz\node[draw,circle,very thick,fill=red, inner sep=0pt,minimum size=7pt]{};) and right (\tikz\node[draw,very thick,fill=blue, inner sep=0pt,minimum size=7pt]{};) resonator determined from the scanning electron micrographs. Scale bar in \bf a \sf and \bf b \sf corresponds to $\SI{1}{\micro\metre}$.
}
\end{figure}
\begin{figure}
\includegraphics[width=\columnwidth/2]{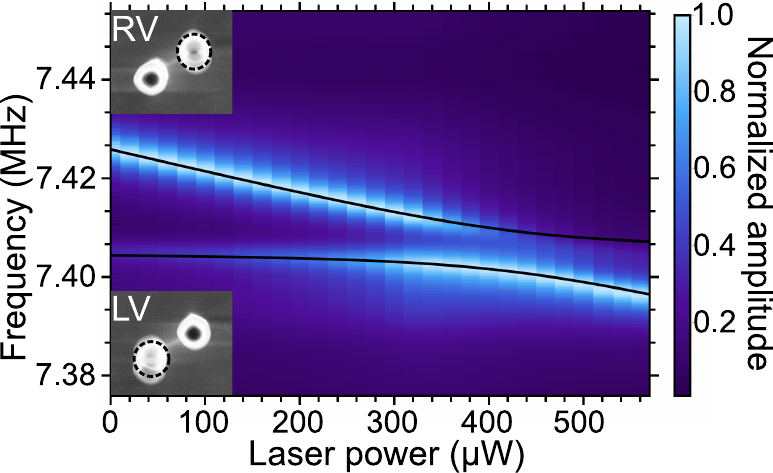}
\caption{\label{Figure3}\bf Strong coupling. \sf Frequency response measurements of the right nanopillar under thermal tuning reveals an avoided crossing as an evidence for strong coupling. The black line shows a fit to the data, yielding a level splitting of $g/2\pi=\SI{8.3}{\kilo\hertz}>\Delta f\approx\SI{3.5}{\kilo\hertz}$, with the linewidth $\Delta f$. Insets show respective scanning electron micrographs of the two modes, indicating a vertical oscillation direction when the resonators are resonantly driven near $f_{\text{LV}}$ (lower inset) and $f_{\text{RV}}$ (upper inset), far from the avoided crossing.
}
\end{figure}
\begin{figure}
\includegraphics[width=\columnwidth/2]{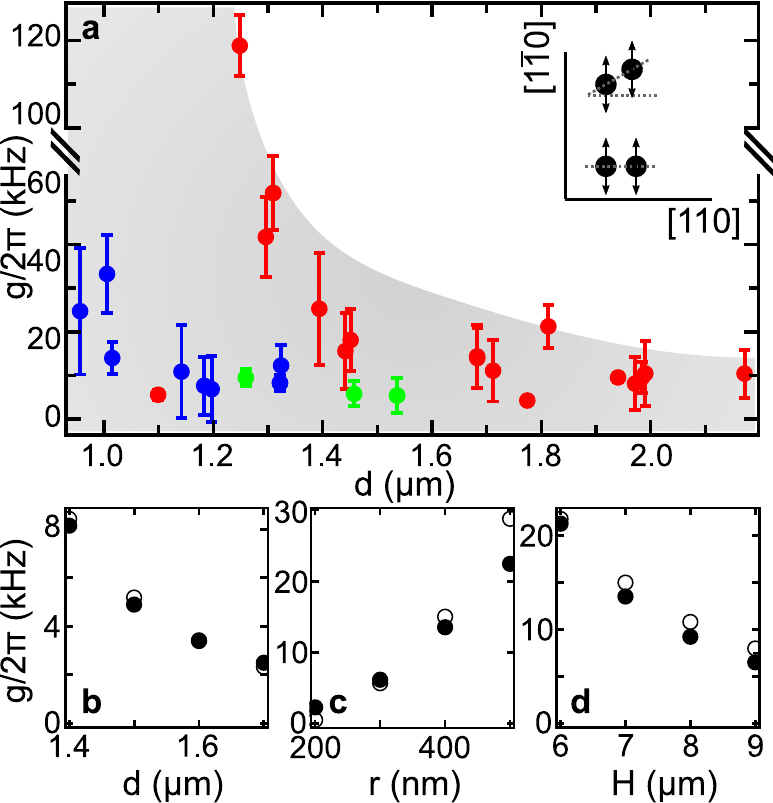}
\caption[]{\label{Figure4}\bf Geometry dependence of coupling strength. a\sf, Experimentally determined coupling rate $g/2\pi$ of the vertical modes of the two nanopillars over their center-to-center distance $d$ for several samples and pillar pair geometries with $r\approx\SI{430}{\nano\metre}$ \& $H\approx\SI{7}{\micro\metre}$ (red), $r\approx\SI{335}{\nano\metre}$ \& $H\approx\SI{7}{\micro\metre}$ (blue), $r\approx\SI{330}{\nano\metre}$ \& $H\approx\SI{8.2}{\micro\metre}$ (green). Error margins describe fitting errors. The inset shows the vibration direction for differently oriented pillar pairs (see Supplementary Information for details). \bf b, c \sf and \bf d\sf, Finite element simulation of level splitting depending on \bf b \sf center-to-center distance $d$ ($r=\SI{400}{\nano\metre}$, $H=\SI{7}{\micro\metre}$), \bf c \sf nanopillar bottom radius $r$ ($d=2r+\SI{400}{\micro\metre}$, $H=\SI{7}{\micro\metre}$) and \bf d \sf nanopillar height $H$ ($r=\SI{400}{\nano\metre}$, $d=\SI{1.2}{\micro\metre}$). We assume an isotropic substrate with Young's Modulus $E_{\text{[100]}}=\SI{85.9}{\giga\pascal}$, and we display vertical (\tikz\node[circle ,fill=black, inner sep=0pt,minimum size=8pt]{};) and horizontal (\tikz\node[draw,circle,very thick,fill=none, inner sep=0pt,minimum size=7pt]{};) modes.
}
\end{figure}

\end{document}